\documentclass[12pt]{article}

\usepackage{times}
\usepackage{graphicx}

\newcounter{lastnote}

\title{Imaging Atomic-Level Random Walk of a Point Defect in Graphene}

\author
{Jani Kotakoski,$^{1,2\ast}$, Clemens Mangler$^{1}$,  Jannik C. Meyer$^{1}$\\
\\
\normalsize{$^{1}$University of Vienna, Faculty of Physics,}\\
\normalsize{Boltzmanngasse 5, A-1090 Vienna, Austria}\\
\normalsize{$^{2}$University of Helsinki, Department of Physics,}\\
\normalsize{P.O. Box 43, FI-00014 Helsinki, Finland}\\
\\
\normalsize{$^\ast$To whom correspondence should be addressed; E-mail: jani.kotakoski@iki.fi.}
}

\date{}

\begin{document} 

\baselineskip24pt

\maketitle 

\section*{Abstract}
{\bf Deviations from the perfect atomic arrangements in crystals play an important
role in affecting their properties. Similarly, diffusion of such deviations is
behind many microstructural changes in solids. However, observation of point
defect diffusion is hindered both by the difficulties related to direct imaging
of non-periodic structures and by the time scales involved in the diffusion
process. Here, instead of imaging thermal diffusion, we stimulate and follow
the migration of a divacancy through graphene lattice using a scanning
transmission electron microscope operated at 60~kV. The beam-activated process
happens on a timescale that allows us to capture a significant part of the
structural transformations and trajectory of the defect.  The low voltage
combined with ultra-high vacuum conditions ensure that the defect remains
stable over long image sequences, which allows us for the first time to
directly follow the diffusion of a point defect in a crystalline material.  }


\section*{Introduction}

High resolution electron microscopy has recently exposed the atomic structure
of two-dimensional materials such as
graphene~\cite{hashimoto_direct_2004,meyer_direct_2008},
hexagonal boron nitride~\cite{jin_fabrication_2009,meyer_selective_2009} and
transition metal di\-chalcogenide~\cite{komsa_two-dimensional_2012} monolayers,
and a two-dimensional silica glass structure~\cite{huang_direct_2012} for
direct observation. Modern imaging techniques are also able to directly discern
between different chemical elements in these structures, even for atoms which
are neighbors in the periodic
table~\cite{krivanek_atom-by-atom_2010,meyer_experimental_2011,zhou_direct_2012,ramasse_probing_2013}.
In addition to structural and chemical analysis, electron microscopy has also
led to advances in the understanding of dynamical, mostly beam-driven,
processes in graphene and similar materials. For example, imaging with 80~keV
electrons has been shown to excite structural changes in pristine
graphene~\cite{meyer_direct_2008}, and at its point
defects~\cite{kotakoski_stone-walestype_2011,robertson_spatial_2012,robertson_structural_2013,wang_direct_2014},
grain boundaries~\cite{kurasch_atom-by-atom_2012} and
dislocations~\cite{warner_dislocation-driven_2012,lehtinen_atomic_2013},
whereas 60~keV imaging has revealed dynamics of a 6-atomic Si
cluster~\cite{lee_direct_2013} in graphene. However, beam-induced knock-on
damage with less-than-ideal vacuum conditions have until the current
state-of-the-art instruments prevented direct observation of point defect
migration in pristine graphene over long image sequences.
Here, using 60~keV imaging, we reveal at the atomic resolution a random walk
performed by a defect in an otherwise perfect graphene crystal.

\section*{Results}

\begin{figure*}[h]
\includegraphics[width=.95\linewidth]{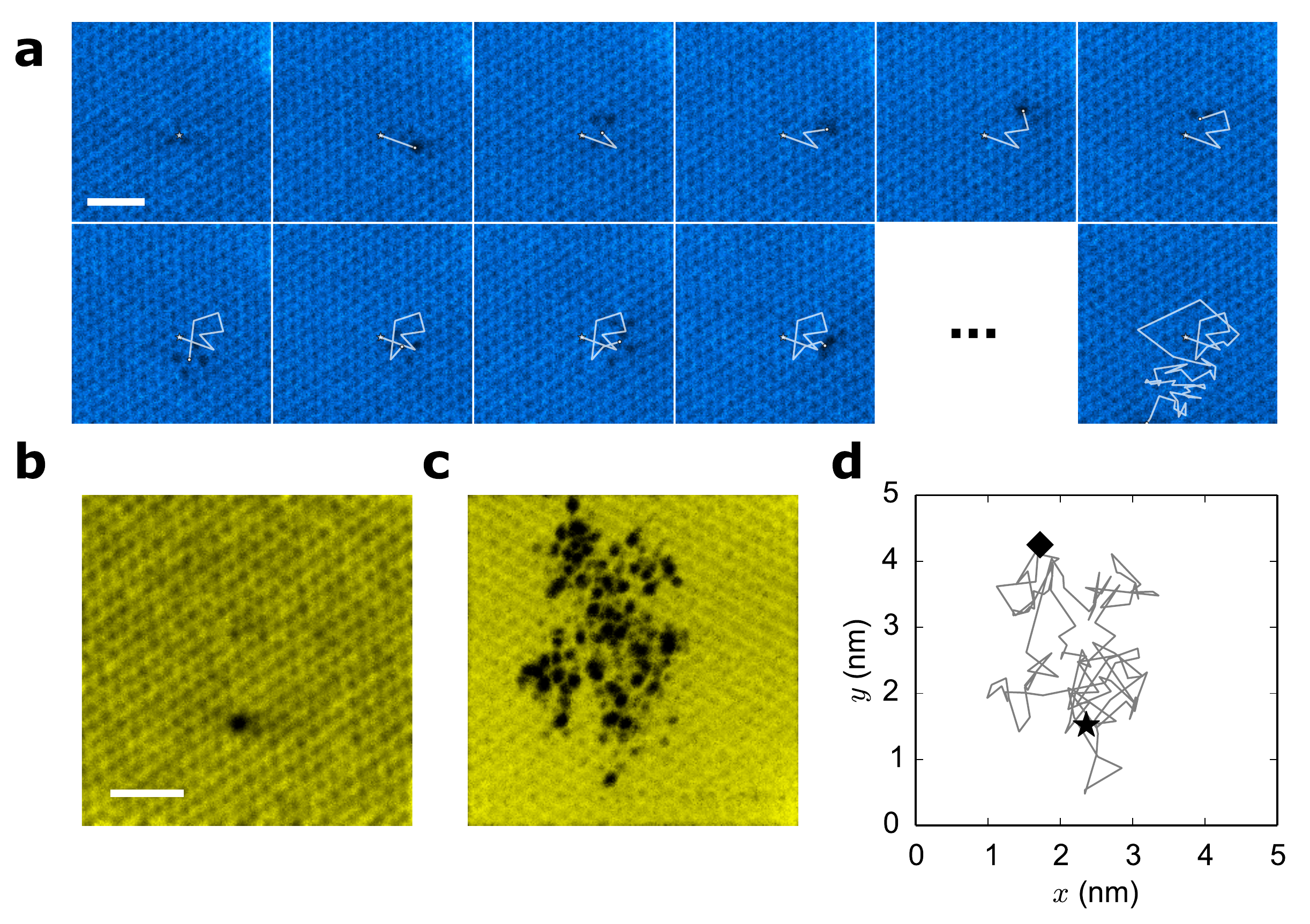}
\caption{{\bf Traveling divacancy in graphene.} (a) Ten consecutive frames and
the final frame from one image sequence showing the movement of the defect
through the lattice. (b) First frame of another image sequence. (c)
"Superposition" of all of the frames from the second sequence highlighting the
trace of the defect by showing the minimum intensity from the sequence at every
pixel. (d) Actual trajectory of the defect in the second sequence, determined
by locating the approximate middle point of the defect in every frame. 
Only those images where the location of the defect was clearly identifiable
have been included. The start position is marked with a black star and the last
location with a diamond.  All scale bars are 1~nm.} \label{fig::traj}
\end{figure*}

\noindent {\bf Image sequences of divacancy migration.} Our data
consist of two images sequences of a divacancy defect in monolayer graphene
(with 57 and 143 frames, recorded over about 5~min 32~s and 12~min 59~s,
respectively), obtained with a Nion UltraSTEM 100~\cite{krivanek_electron_2008}
electron microscope operated at 60~kV.  Due
to the low voltage and the ultra-high vacuum conditions ($1.3\times
10^{-9}$~mbar at the sample), the divacancy in graphene is extraordinarily
stable: it neither converts into higher-order vacancies nor traps carbon and
converts back to a pristine lattice, for long sequences of images. However, the
defect rapidly moves via beam-driven bond rotations during observation, and
constantly changes shape between four different configurations, which have also
been identified in earlier images~\cite{banhart_structural_2011}. As a result,
the defect performs a random walk through the lattice.  A sequence of 10
consecutive frames and a final frame of one image sequence is presented in
Fig.~\ref{fig::traj}a as an example. Fig.~\ref{fig::traj}b shows the first
frame from another image sequence. Fig.~\ref{fig::traj}c is a ``superposition''
of all frames highlighting the trace of the defect (by showing the minimum of
intensity from the sequence at every pixel), and Fig.~\ref{fig::traj}d depicts
the defect trajectory from the same sequence, obtained by locating the center
of the divacancy in every frame. Complete sequences are provided as
supplementary movie 1 for the first image sequence and as supplementary movie 2
for the second image sequence. Non-treated versions of all images are provided
as supplementary movie 3 and 4 for the first and the second sequence,
respectively.

\begin{figure*}[h]
\includegraphics[width=\linewidth]{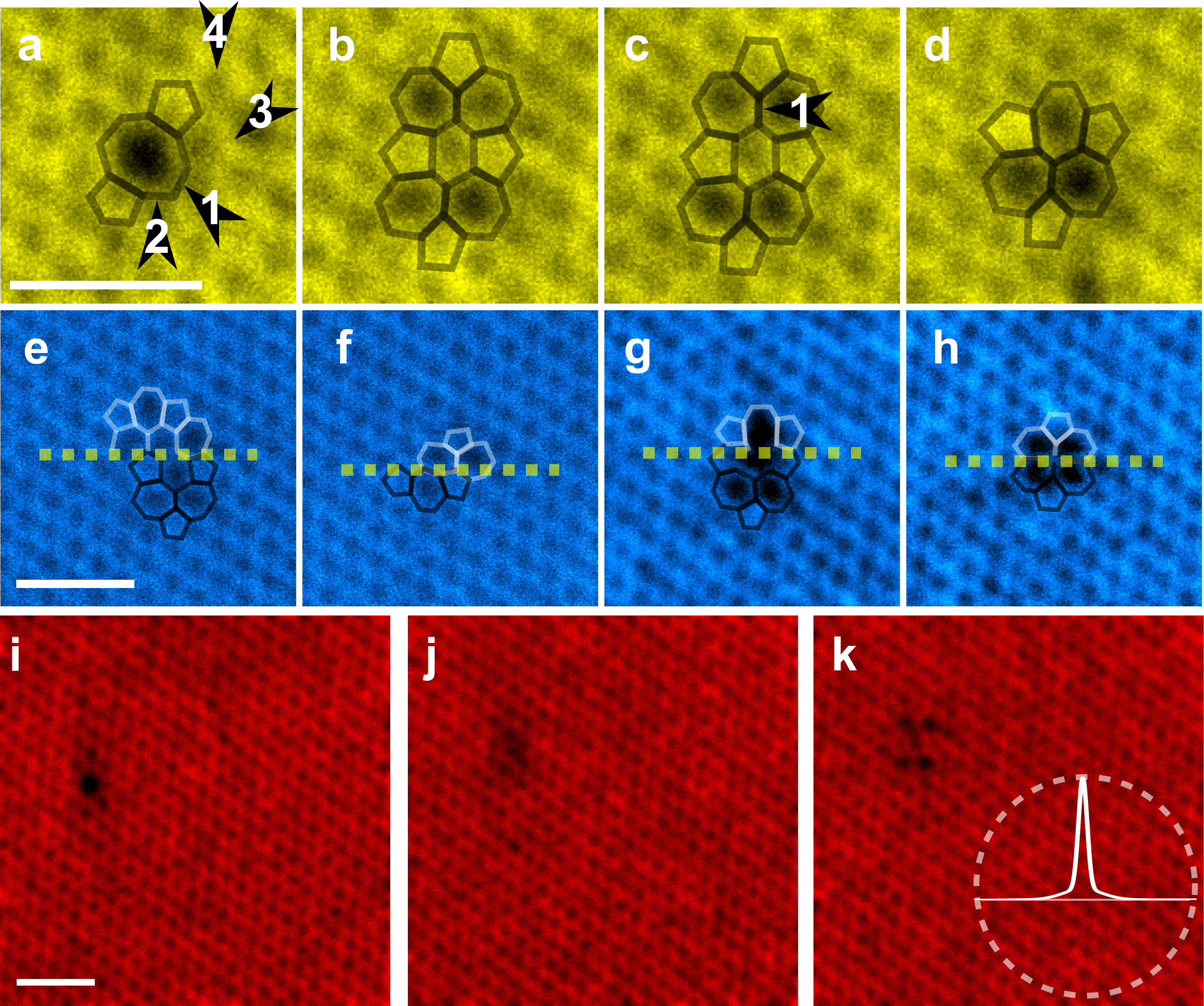} 
\caption{{\bf Example exposures
of the defect.} (a-d) Four subsequent frames from the second image sequence.
The bonds associated with the defects are highlighted with an overlay. The
structure has undergone at least four bond rotations between panels a and b and
one between c and d, as marked with the black arrows. (e--h) Examples of image
scans where the structure changed while scanning exactly at the location of the
defect. White and black overlays mark the structure of the defect before and
after the change, respectively. (i--k) Three subsequent frames from the second
image sequence.  The defect appears in the V$_2$(585) configuration in panel
i, but disappears for the duration of the next scan resulting in panel j,
before appearing again in panel k, in the V$_2$(5555-6-7777) configuration. The
darker area within panel j presumably corresponds to the area where the
defect is located, although it avoids detection (locations of all atoms
belonging to the pristine lattice can be identified). In panel k, a circle with
radius of 1.5~nm is drawn for scale with the experimentally obtained probe
shape. In these panels the complete field of view is shown. The scale bars are
1~nm.} \label{fig::defs} \end{figure*}

\noindent {\bf Atomic configurations of the divacancy.} Fig.~\ref{fig::defs} a--d shows
the defect in three different configurations: V$_2$(585) in panel a,
V$_2$(5555-6-7777) in panels b and c, and V$_2$(555-777) in panel d. (In this
notation, the carbon rings contained within the defects are listed, when
possible, along the longest axis through the defect.) Due to the symmetry of
the lattice, V$_2$(555-777) can appear with two distinct orientations, whereas
the other two have three possible conformations.  Throughout the data, these
defects have occurrancies of 50.3\% for V$_2$(585), 14.1\% for  V$_2$(555-777)
and 18.8\% for V$_2$(5555-6-7777). Additionally, in three frames (1.8\%) the
defect appears in the $2\times (57)$ configuration~\cite{kotakoski_point_2011}.
The rest (14.7\%) of the frames contain either unclear structures or
combinations of two or more of the above-mentioned configurations. Examples of
these are shown in Fig.~\ref{fig::defs} e--h. In two frames the defect has
completely eluded detection although it is visible both in the previous and the
following frames (an example is presented in Fig.~\ref{fig::defs}i-k).

\noindent {\bf Defect transformations.} As mentioned, less than 15\% of the
scan images showed any indications of the structure changing during the scan.
This is surprising, because the consecutive scans nevertheless revealed the
defect in other locations and often in another configuration. It is almost as
if the scans would correspond to photographs taken in a busy but dark room,
only momentarily illuminated by the flash of the camera freezing the moment in
time. However, it is well understood that the bond rotations, which are
responsible for the observed structural changes, are associated with an energy
barrier of 5--10~eV~\cite{li_defect_2005} and can thus not be driven by thermal
activation in our room temperature experiments. Indeed, they must be caused by
collisions between individual imaging electrons and individual target
nuclei~\cite{kotakoski_stone-walestype_2011}. Therefore, one could expect
that the transformations always occur when the electron probe is placed atop
the defect, which would necessarily lead to detection of the transformation, as
in Fig.~\ref{fig::defs}e-h. Perhaps even more mysterious are the two frames
where the defect has completely avoided detection (for an example, see
Fig.~\ref{fig::defs}i-k).

\begin{figure}[h]
\includegraphics[width=\linewidth]{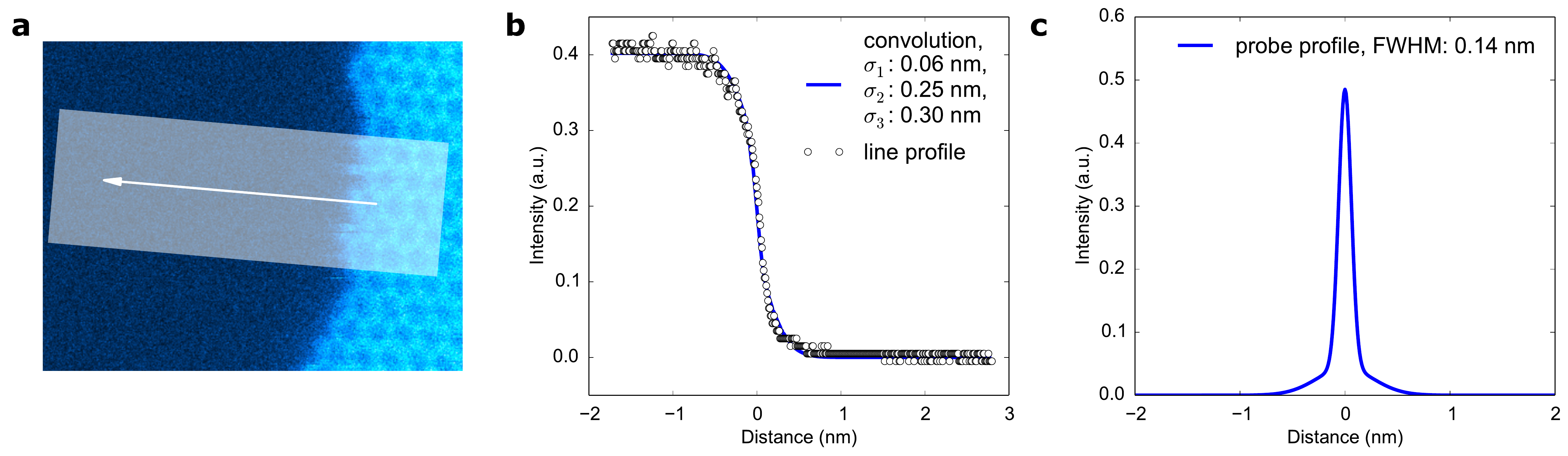} 
\caption{{\bf Determination of the
probe shape.} (a) Unprocessed (but colored) image of a graphene edge. (b) Line
profile obtained from the area shaded in panel a in the direction of the arrow
along with a simulated line profile calculated assuming that the graphene edge
is a step function and convoluting it with a probe that consists of three 2D
Gaussians. The graphene edge position was set to the zero of $x$-axis. The
standard deviations for the Gaussians ($\sigma_1$, $\sigma_2$ and $\sigma_3$)
were obtained via manual fitting, with an estimated accuracy of
ca. 10\%. (c) One dimensional profile of the probe consisting of the three
Gaussians. Vacuum intensity was set to zero.}\label{fig::probe}
\end{figure}

\noindent {\bf Determination of the probe shape.} The answer to the
transformation puzzle lies in the shape of the electron probe, which is pixel
by pixel and line by line scanned over the area within the field of view (in
our case $512\times 512$ pixels within $4\times 4$~nm$^2$ for sequence 1 and
$5\times 5$~nm$^2$ for sequence 2). While the full width at half maximum
(FWHM) of the probe must be in the order of 1~{\AA} for
atomic resolution imaging, the actual shape of the probe, and especially its
tail~\cite{krivanek_atom-by-atom_2010} further away from the point of the
maximum intensity is not exactly known. In order to understand the role of the probe
tails in our observations, we estimated the shape of our probe experimentally
based on an intensity profile recorded over a graphene edge, as illustrated in
Fig.~\ref{fig::probe}.  A good match between the recorded profile and a
convolution of a step function with a model probe consisting of three 2D
Gaussians (see Fig.~\ref{fig::probe}b) was obtained for standard deviations of
$\sigma_1 \approx 0.06$~nm, $\sigma_2 \approx 0.25$~nm and $\sigma_3 \approx
0.30$~nm.  The estimated accuracy of the manual fit is ca. 10\%. A
one-dimensional profile of the measured probe is presented in
Fig.~\ref{fig::probe}c. The FWHM for the probe is about
0.14~nm and the vacuum level is reached at a distance of about 1.5~nm.  Less
than 21\% of the beam intensity is contained within the FWHM, which shows that
a significant dose is deposited outside the beam maximum position.  Based on
this, and taking into account that the probe spends much more time at a
distance of $0.07~\mathrm{nm}< r \leq 1.5$~nm from the defect than on the defect itself [the middle
sized defect V$_2$(555-777) has an area roughly 5\% of that of $\pi
(1.5~\mathrm{nm})^2$], it becomes clear that the effect of scanning near the
defect can alter its atomic configuration. As noted, in our experiment this
effect accounted for up to $>85$\% of all of the transformations. When the
transformations happen to drive the defect towards the already imaged area,
this effect leads to its apparent disappearance for the duration of one or more scans.

\begin{figure*}[h]
\includegraphics[width=\linewidth]{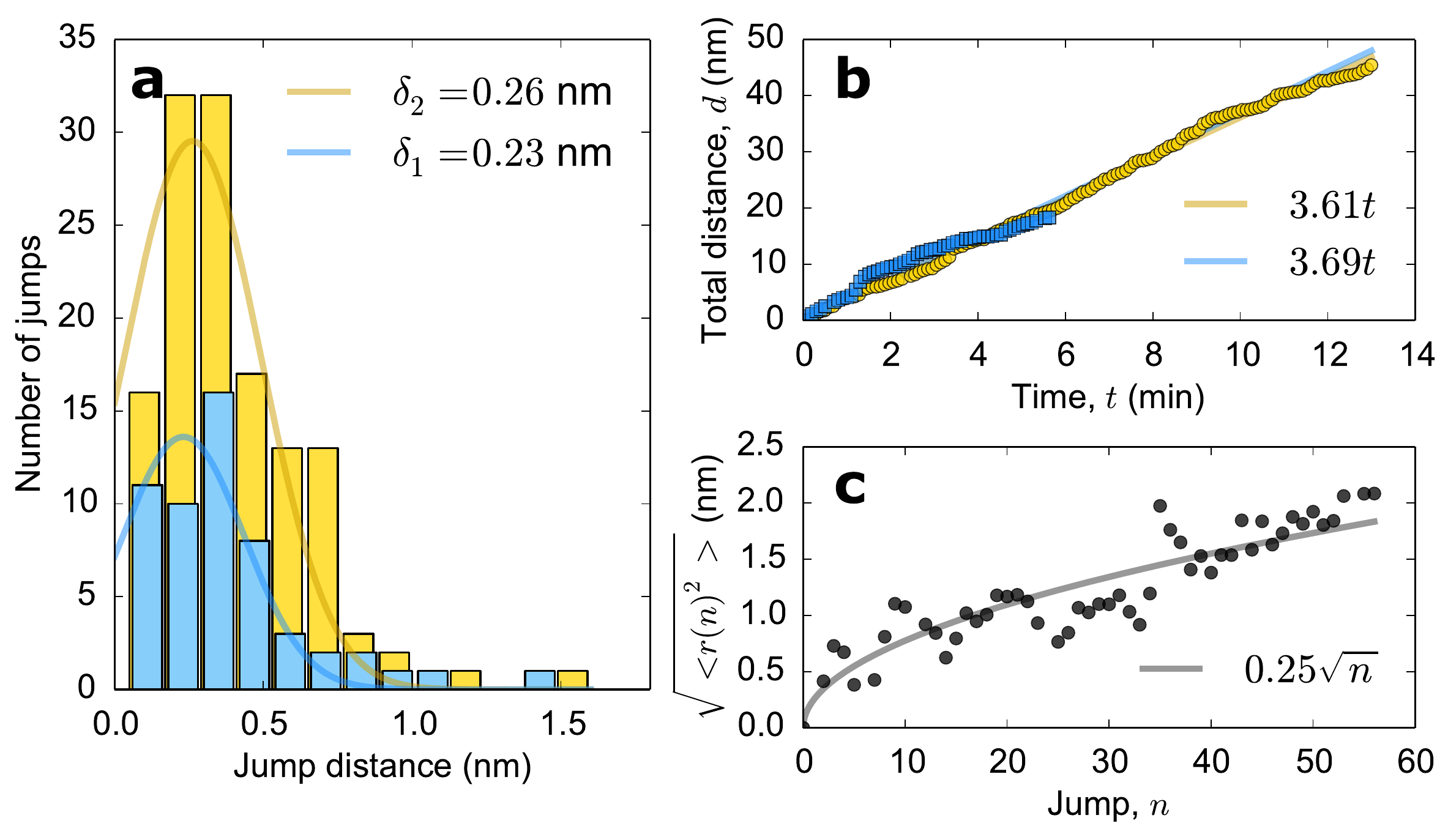}
\caption{{\bf Statistical analysis of a random walk.} (a) Histogram of all of
the jump distances by the defect in the two image sequences.  The lines show
normal distributions fitted to the data. (b) Corresponding cumulative total
distance traveled by the defect as a function of time. The lines are fits to
the data. (c) Root-mean-square distance of the defect from the starting
position as a function of time. The solid line is a fit to the data.}
\label{fig::stats} \end{figure*}

\noindent {\bf Statistical analysis of the random walks.} Histograms of all of the jumps
between recorded frames in both of the image sequences are plotted in
Fig.~\ref{fig::stats}a. Assuming normal distribution, the average jump length
is $\delta_1 \approx 0.23$~nm for sequence 1 and $\delta_2 \approx 0.26$~nm for
sequence 2, which is close to the hexagon-hexagon distance in the carbon
lattice. From the definition of diffusivity in two dimensions
\begin{equation}
  D \equiv \frac{\delta^2}{4\tau},
\end{equation}
where $\tau$ is the jump time, we get
$D \approx 3.10\times 10^{-3}~\mathrm{nm}^2\mathrm{s}^{-1} = 3.1\times
10^{-21}$~m$^2\mathrm{s}^{-1}$ (using the value for the longer sequence), which is well in
line with typical diffusivity values in solids. However, we stress that the
migration is in the case of our experiment driven by the knock-on collisions
between individual electrons and individual target nuclei, and the measured
diffusivity is thus not directly comparable to values describing thermally
driven point defect diffusion in solids.

Although some of the jumps between scans are considerably longer than others,
the overall total cumulative distance ($d$) traveled by the defect increases
linearly as a function of time (see Fig.~\ref{fig::stats}b), yielding an
estimated migration speed of 3.61~nm~min$^{-1}$ during the 13~min long experiment. 
For a random walk, the root-mean-square distance, which is a measure of the
average distance of the walker from the start after $n$ steps, is
\begin{equation}
  \sqrt{<r(n)^2>} \equiv \sqrt{\frac{1}{N}\sum_{k=1}^N r_k(n)^2} = \delta\sqrt{n},
\end{equation}
where $k$ runs over the $N$ different random walks (in our case the two image
sequences) and $\delta$ is the average jump length. As can be seen in
Fig.~\ref{fig::stats}c, the random walks analyzed here follow this behaviour. A
fit to the data reveals $\delta \approx 0.25$~nm, which is in between the
above-estimated values of $\delta_1$ and $\delta_2$, as can be expected.

We can take the analogue between thermal and electron-beam-driven diffusion one
step further to compare our experimental observations with surface diffusion,
where
\begin{equation}
  \Gamma(T) = \nu \exp(-E_b/k_BT).
\end{equation}
Here $\Gamma (T)$ is the jump rate at temperature $T$, $\nu$ the
attempt frequency, $E_b$ the diffusion energy barrier, and $k_B$ the Boltzmann
constant. For the area of the middle-sized divacancy, V$_2$(555-777), and $E_b
= 5$~eV as the energy barrier associated with a bond
rotation~\cite{li_defect_2005}, we get an estimate for the conditions which we
simulate with the electron beam. The apparent temperature of the system is $T
\approx 3050$~K, with the caveat that every recorded image is here assumed to
represent exactly one migration step, whereas in reality we
know that often several bond rotations have taken place between two subsequent
frames. We stress that this is a virtual temperature, since the actual heat
brought in by the electron beam is quickly dissipated away, and only modest if
any actual heating of the sample is expected during the
experiment~\cite{egerton_radiation_2004}. Nevertheless, the observed defect
migration is otherwise indistinquishable from what would be expected to occur
at elevated temperatures: The energy input from the beam helps to overcome an
activation barrier; in a similar manner as heat would
but---importantly---resulting in a much slower process.

\section*{Discussion}

As a conclusion, we have demonstrated for the first time that electron
irradiation at 60 kV can be used to stimulate an atomic-scale random walk of a
point defect in an otherwise pristine crystal, and to record it over several
minutes.  Despite the atomically small probe size, most of the transformations
occur when the beam is situated away from the actual defect due to irradiation
dose accumulation from the low-intensity tail of the electron beam. In rare
cases, the defect can even completely avoid detection during a scan. Via
analysis of the defect trajectory during the experiment, we estimate that the
beam-stimulated migration of the divacancy corresponds to a virtual temperature
of about 3050~K, establishing atomic resolution transmission electron
microscopy as a method for simultaneous imaging and driving diffusion of point
defects in low-dimensional materials.

\section*{Methods}

\noindent {\bf Scanning transmission electron microscopy.} The experiments were carried
out with a Nion UltraSTEM 100 device~\cite{krivanek_electron_2008} recently
installed at the University of Vienna. The device is equipped with a cold field
emission gun, which was operated at 60~kV in ultrahigh vacuum. Medium angle
annular dark-field detector was used to record two image sequences of the same
defect. The first sequence contains 57 frames recorded over about 5~min and
32~s with a field of view of $4\times 4$~nm$^2$. The second sequence contains
143 frames and was recorded over about 12~min and 59~s. Sample drift is less
than the lattice spacing during the image sequences (the image sequences
provided in supplementary movies 1--4 were not compensated for drift). The time
the beam was held at each pixel was 16~$\mu$s for both sequences and all
images contain $512\times 512$ pixels.  A typical camera current for the device
is in the order of $5\times 10^{-11}$~A, from which a dose of circa $8\times
10^{6} e^-${\AA}$^{-2}$ was estimated per recorded frame for sequence 1 and
$5\times 10^{6} e^-${\AA}$^{-2}$
for sequence 2.

\noindent {\bf Image processing.} The images were processed to reduce noise by applying a
Gaussian filter with a radius of 6~px (for field of view of $4\times 4$~nm$^2$)
and 4~px (for field of view of $5\times 5$~nm$^2$) after which the processed
image was multiplied with the original image. The process was carried out twice
for the second image sequence due to lower signal-to-noise ratio.

\noindent {\bf Determination of the defect position.} To track the migration of
the defect, we first aligned the image sequences, and then marked the middle of
the defect manually to obtain the coordinates for each frame. The accuracy of
this positioning procedure is estimated to be better than the interatomic
separation in the lattice.

\noindent {\bf Sample preparation.} The graphene sample was grown via chemical vapor
deposition and suspended on a TEM grid by a commercial supplier.

\section*{Acknowledgments}
We acknowledge funding from the Austrian Science
Fund (FWF) through projects M~1481-N20 and P~25721-N20 and from University of Helsinki
Funds. 

\section*{Author contributions}
J.K. initiated the study, carried out the experiment, analyzed data and
wrote the manuscript. C.M. contributed to establishing the experimental setup and
J.C.M. contributed to the analysis. All authors commented on the manuscript.

\section*{Additional information}
{\bf Supplementary Information:} accompanies this paper at http://www.nature.com/naturecommunications

\noindent{\bf Competing financial interests:} The authors declare no competing financial interests.

\newpage
\clearpage 

\section*{Figures}

\begin{thebibliography}{10}
\expandafter\ifx\csname url\endcsname\relax
  \def\url#1{\texttt{#1}}\fi
\expandafter\ifx\csname urlprefix\endcsname\relax\def\urlprefix{URL }\fi
\providecommand{\bibinfo}[2]{#2}
\providecommand{\eprint}[2][]{\url{#2}}

\bibitem{hashimoto_direct_2004}
\bibinfo{author}{Hashimoto, A.}, \bibinfo{author}{Suenaga, K.},
  \bibinfo{author}{Gloter, A.}, \bibinfo{author}{Urita, K.} \&
  \bibinfo{author}{Iijima, S.}
\newblock \bibinfo{title}{Direct evidence for atomic defects in graphene
  layers}.
\newblock \emph{\bibinfo{journal}{Nature}} \textbf{\bibinfo{volume}{430}},
  \bibinfo{pages}{870--873} (\bibinfo{year}{2004}).

\bibitem{meyer_direct_2008}
\bibinfo{author}{Meyer, J.~C.} \emph{et~al.}
\newblock \bibinfo{title}{Direct imaging of lattice atoms and topological
  defects in graphene membranes.}
\newblock \emph{\bibinfo{journal}{Nano Lett.}} \textbf{\bibinfo{volume}{8}},
  \bibinfo{pages}{3582--3586} (\bibinfo{year}{2008}).

\bibitem{jin_fabrication_2009}
\bibinfo{author}{Jin, C.}, \bibinfo{author}{Lin, F.}, \bibinfo{author}{Suenaga,
  K.} \& \bibinfo{author}{Iijima, S.}
\newblock \bibinfo{title}{Fabrication of a freestanding boron nitride single
  layer and its defect assignments}.
\newblock \emph{\bibinfo{journal}{Phys. Rev. Lett.}}
  \textbf{\bibinfo{volume}{102}}, \bibinfo{pages}{195505}
  (\bibinfo{year}{2009}).

\bibitem{meyer_selective_2009}
\bibinfo{author}{Meyer, J.~C.}, \bibinfo{author}{Chuvilin, A.},
  \bibinfo{author}{Algara-Siller, G.}, \bibinfo{author}{Biskupek, J.} \&
  \bibinfo{author}{Kaiser, U.}
\newblock \bibinfo{title}{Selective sputtering and atomic resolution imaging of
  atomically thin boron nitride membranes.}
\newblock \emph{\bibinfo{journal}{Nano Lett.}} \textbf{\bibinfo{volume}{9}},
  \bibinfo{pages}{2683--2689} (\bibinfo{year}{2009}).

\bibitem{komsa_two-dimensional_2012}
\bibinfo{author}{Komsa, H.-P.} \emph{et~al.}
\newblock \bibinfo{title}{Two-dimensional transition metal dichalcogenides
  under electron irradiation: defect production and doping}.
\newblock \emph{\bibinfo{journal}{Phys. Rev. Lett.}}
  \textbf{\bibinfo{volume}{109}}, \bibinfo{pages}{035503}
  (\bibinfo{year}{2012}).

\bibitem{huang_direct_2012}
\bibinfo{author}{Huang, P.~Y.} \emph{et~al.}
\newblock \bibinfo{title}{Direct imaging of a two-dimensional silica glass on
  graphene}.
\newblock \emph{\bibinfo{journal}{Nano Lett.}} \textbf{\bibinfo{volume}{12}},
  \bibinfo{pages}{1081--1086} (\bibinfo{year}{2012}).

\bibitem{krivanek_atom-by-atom_2010}
\bibinfo{author}{Krivanek, O.~L.} \emph{et~al.}
\newblock \bibinfo{title}{Atom-by-atom structural and chemical analysis by
  annular dark-field electron microscopy.}
\newblock \emph{\bibinfo{journal}{Nature}} \textbf{\bibinfo{volume}{464}},
  \bibinfo{pages}{571--574} (\bibinfo{year}{2010}).

\bibitem{meyer_experimental_2011}
\bibinfo{author}{Meyer, J.~C.} \emph{et~al.}
\newblock \bibinfo{title}{Experimental analysis of charge redistribution due to
  chemical bonding by high-resolution transmission electron microscopy.}
\newblock \emph{\bibinfo{journal}{Nature Mater.}}
  \textbf{\bibinfo{volume}{10}}, \bibinfo{pages}{209--215}
  (\bibinfo{year}{2011}).

\bibitem{zhou_direct_2012}
\bibinfo{author}{Zhou, W.} \emph{et~al.}
\newblock \bibinfo{title}{Direct determination of the chemical bonding of
  individual impurities in graphene}.
\newblock \emph{\bibinfo{journal}{Phys. Rev. Lett.}}
  \textbf{\bibinfo{volume}{109}}, \bibinfo{pages}{206803}
  (\bibinfo{year}{2012}).

\bibitem{ramasse_probing_2013}
\bibinfo{author}{Ramasse, Q.~M.} \emph{et~al.}
\newblock \bibinfo{title}{Probing the bonding and electronic structure of
  single atom dopants in graphene with electron energy loss spectroscopy}.
\newblock \emph{\bibinfo{journal}{Nano Letters}} \textbf{\bibinfo{volume}{13}},
  \bibinfo{pages}{4989--4995} (\bibinfo{year}{2013}).

\bibitem{kotakoski_stone-walestype_2011}
\bibinfo{author}{Kotakoski, J.} \emph{et~al.}
\newblock \bibinfo{title}{Stone-wales–type transformations in carbon
  nanostructures driven by electron irradiation}.
\newblock \emph{\bibinfo{journal}{Phys. Rev. B}} \textbf{\bibinfo{volume}{83}},
  \bibinfo{pages}{245420} (\bibinfo{year}{2011}).

\bibitem{robertson_spatial_2012}
\bibinfo{author}{Robertson, A.~W.} \emph{et~al.}
\newblock \bibinfo{title}{Spatial control of defect creation in graphene at the
  nanoscale}.
\newblock \emph{\bibinfo{journal}{Nature Comm.}} \textbf{\bibinfo{volume}{3}},
  \bibinfo{pages}{1144} (\bibinfo{year}{2012}).

\bibitem{robertson_structural_2013}
\bibinfo{author}{Robertson, A.~W.} \emph{et~al.}
\newblock \bibinfo{title}{Structural reconstruction of the graphene
  monovacancy}.
\newblock \emph{\bibinfo{journal}{{ACS} Nano}} \textbf{\bibinfo{volume}{7}},
  \bibinfo{pages}{4495--4502} (\bibinfo{year}{2013}).

\bibitem{wang_direct_2014}
\bibinfo{author}{Wang, W.~L.} \emph{et~al.}
\newblock \bibinfo{title}{Direct observation of a long-lived single-atom
  catalyst chiseling atomic structures in graphene}.
\newblock \emph{\bibinfo{journal}{Nano Letters}} \textbf{\bibinfo{volume}{14}},
  \bibinfo{pages}{450--455} (\bibinfo{year}{2014}).

\bibitem{kurasch_atom-by-atom_2012}
\bibinfo{author}{Kurasch, S.} \emph{et~al.}
\newblock \bibinfo{title}{Atom-by-atom observation of grain boundary migration
  in graphene}.
\newblock \emph{\bibinfo{journal}{Nano Lett.}} \textbf{\bibinfo{volume}{12}},
  \bibinfo{pages}{3168--3173} (\bibinfo{year}{2012}).

\bibitem{warner_dislocation-driven_2012}
\bibinfo{author}{Warner, J.~H.} \emph{et~al.}
\newblock \bibinfo{title}{Dislocation-driven deformations in graphene}.
\newblock \emph{\bibinfo{journal}{Science}} \textbf{\bibinfo{volume}{337}},
  \bibinfo{pages}{209--212} (\bibinfo{year}{2012}).

\bibitem{lehtinen_atomic_2013}
\bibinfo{author}{Lehtinen, O.}, \bibinfo{author}{Kurasch, S.},
  \bibinfo{author}{Krasheninnikov, A.~V.} \& \bibinfo{author}{Kaiser, U.}
\newblock \bibinfo{title}{Atomic scale study of the life cycle of a dislocation
  in graphene from birth to annihilation}.
\newblock \emph{\bibinfo{journal}{Nature Comm.}} \textbf{\bibinfo{volume}{4}},
  \bibinfo{pages}{2098} (\bibinfo{year}{2013}).

\bibitem{lee_direct_2013}
\bibinfo{author}{Lee, J.}, \bibinfo{author}{Zhou, W.},
  \bibinfo{author}{Pennycook, S.~J.}, \bibinfo{author}{Idrobo, J.-C.} \&
  \bibinfo{author}{Pantelides, S.~T.}
\newblock \bibinfo{title}{Direct visualization of reversible dynamics in a si6
  cluster embedded in a graphene pore}.
\newblock \emph{\bibinfo{journal}{Nature Comm.}} \textbf{\bibinfo{volume}{4}},
  \bibinfo{pages}{1650} (\bibinfo{year}{2013}).

\bibitem{krivanek_electron_2008}
\bibinfo{author}{Krivanek, O.~L.} \emph{et~al.}
\newblock \bibinfo{title}{An electron microscope for the aberration-corrected
  era}.
\newblock \emph{\bibinfo{journal}{Ultramicroscopy}}
  \textbf{\bibinfo{volume}{108}}, \bibinfo{pages}{179--195}
  (\bibinfo{year}{2008}).

\bibitem{banhart_structural_2011}
\bibinfo{author}{Banhart, F.}, \bibinfo{author}{Kotakoski, J.} \&
  \bibinfo{author}{Krasheninnikov, A.~V.}
\newblock \bibinfo{title}{Structural defects in graphene.}
\newblock \emph{\bibinfo{journal}{{ACS} Nano}} \textbf{\bibinfo{volume}{5}},
  \bibinfo{pages}{26--41} (\bibinfo{year}{2011}).

\bibitem{kotakoski_point_2011}
\bibinfo{author}{Kotakoski, J.}, \bibinfo{author}{Krasheninnikov, A.~V.},
  \bibinfo{author}{Kaiser, U.} \& \bibinfo{author}{Meyer, J.~C.}
\newblock \bibinfo{title}{From point defects in graphene to two-dimensional
  amorphous carbon}.
\newblock \emph{\bibinfo{journal}{Phys. Rev. Lett.}}
  \textbf{\bibinfo{volume}{106}}, \bibinfo{pages}{105505}
  (\bibinfo{year}{2011}).

\bibitem{li_defect_2005}
\bibinfo{author}{Li, L.}, \bibinfo{author}{Reich, S.} \&
  \bibinfo{author}{Robertson, J.}
\newblock \bibinfo{title}{Defect energies of graphite: Density-functional
  calculations}.
\newblock \emph{\bibinfo{journal}{Phys. Rev. B}} \textbf{\bibinfo{volume}{72}},
  \bibinfo{pages}{184109} (\bibinfo{year}{2005}).

\bibitem{egerton_radiation_2004}
\bibinfo{author}{Egerton, R.}, \bibinfo{author}{Li, P.} \&
  \bibinfo{author}{Malac, M.}
\newblock \bibinfo{title}{Radiation damage in the {TEM} and {SEM}}.
\newblock \emph{\bibinfo{journal}{Micron}} \textbf{\bibinfo{volume}{35}},
  \bibinfo{pages}{399--409} (\bibinfo{year}{2004}).

\end{thebibliography}
\end{document}